\documentclass[12pt]{article} 
\usepackage{amssymb,amsmath,amsfonts} 
\usepackage[dvips]{graphicx}

\begin{document} 
 
\title{Crossing probabilities on same-spin clusters\\
in the two-dimensional Ising model} 
\author{ 
Ervig Lapalme\footnote{Centre de recherches math\'ematiques, 
Universit\'e de Montr\'eal, C.P.\ 6128, succ.\ centre-ville,
Montr\'eal,  Qu\'ebec, Canada\ \ H3C 3J7. Email: {\ttfamily
lapalmee{\char'100}crm.umontreal.ca}}
\and Yvan Saint-Aubin\footnote{
Centre de recherches math\'ematiques and 
D\'ept.\ de math\'ematiques et de statistique,
Universit\'e de Montr\'eal, C.P.\ 6128, succ.\ centre-ville,
Montr\'eal,  Qu\'ebec, Canada\ \ H3C 3J7.
Email: {\ttfamily saint{\char'100}crm.umontreal.ca}}
}
\date{\today}

\maketitle 
 
\begin{abstract} 
Probabilities of crossing on same-spin clusters, seen as order
parameters, have been introduced recently for the critical 2d Ising model by
Langlands, Lewis and Saint-Aubin.
We extend Cardy's ideas, introduced for percolation, to obtain an
ordinary  differential equation of order $6$ for the horizontal crossing probability
$\pi_h$. Due to the identity $\pi_h(r)+\pi_h(1/r)=1$, the function $\pi_h$
must lie in a 3-dimensional subspace. New measurements of $\pi_h$ are made
for $40$ values of the aspect ratio
$r$ ($r\in[0.1443,6.928]$). These data are more precise than those obtained
by Langlands {\em et al} as
the $95$\%-confidence interval is brought to $4\times10^{-4}$. A $3$-parameter
fit using these new data determines the solution of the differential equation.
The largest gap between this solution and the 40 data is smaller than 
$4\times 10^{-4}$. The probability $\pi_{hv}$ of simultaneous horizontal
and vertical crossings is also treated.

\bigskip

Les probabilit\'es de travers\'ee sur les plages de spins identiques,
vues comme para\-m\`e\-tres d'ordre, ont \'et\'e introduites 
r\'ecemment pour le mod\`ele d'Ising bidimensionnel critique par
Langlands, Lewis et Saint-Aubin. Nous \'etendons
les id\'ees de Cardy, pr\'esent\'ees pour la percolation, afin d'obtenir
une \'equation diff\'erentielle ordinaire du sixi\`eme ordre pour la
probabilit\'e d'une travers\'ee horizontale $\pi_h$. Il suit de 
l'identit\'e $\pi_h(r)+\pi_h(1/r)=1$ que $\pi_h$ doit reposer dans un 
sous-espace de dimension trois. De nouvelles mesures de $\pi_h$ sont faites
pour $40$ valeurs du rapport $r$ des c\^ot\'es du rectangle ($r\in[0.1443,6.928]$).
L'intervalle de confiance de $95$\% est ici r\'eduit \`a $4\times 10^{-4}$.
La solution de l'\'equation diff\'erentielle est choisie en minimisant
l'\'ecart quadratique entre la solution et ces nouvelle donn\'ees. Le
plus grand \'ecart entre les donn\'ees et la solution ainsi obtenue est
plus petit que $4\times 10^{-4}$. La probabilit\'e de travers\'ees
horizontale et verticale simultan\'ees est \'egalement \'etudi\'ee.
\end{abstract} 

\noindent\hskip\leftmargin{\scshape Short Title:}{} Crossing probabilities in the 2d Ising model

\noindent\hskip\leftmargin{\scshape PACS numbers:}{} 0.50.+q, 64.60.Fr, 64.60.Ak, 11.25.Hf

\section{Introduction} 

The probability $\pi_h(r)$ of crossing on open sites inside a rectangle of aspect
ratio $r$ has been measured at $p_c$ for several models of percolation in \cite{LPPS} and
\cite{LPS}. These simulations support hypotheses of universality and conformal
invariance of this function $\pi_h(r)$ and of several others. Cardy's 
contemporaneous work \cite{Cardyperco} offered a prediction for $\pi_h$ using
conformal field theory. His analytic expression agrees with the simulations
within statistical errors and provides further support that these crossing
probabilities are order parameters with the usual critical properties.

This paper presents similar evidence, both numerical and analytic, for 
crossing probabilities
on same-spin clusters of the two-dimensional Ising model at
criticality. These crossing probabilities are not traditional order parameters
for the Ising model. It is not a priori obvious that they are not identically 
zero or one in any dimension $d$. In dimension two however the simulations carried out
in \cite{LLS} indicates clearly that they are non-trivial functions and that they are
likely to satisfy the same hypotheses (or even more restrictive ones) of 
universality and conformal invariance. Moreover physical quantities for which conformal
field theory gives quantitative predictions that can be readily verified by
simulation are always welcome.

Let a triangular lattice be oriented in such a way that sites on horizontal
lines are at a distance of one mesh unit. A rectangle of height $V$ and width
$H$ is superimposed on the lattice. The width $H$ is measured in mesh units
but $V$ is the number of horizontal lines in the rectangle. A configuration of the 
Ising model has a horizontal crossing if there exists a path made of edges between 
nearest neighbor sites going
from the leftmost inner column to the rightmost one and visiting only plus
spins. (Because of the relative position of the rectangle and the lattice,
the vertical columns are made of sites in zigzag.) Let $\pi_h(V,H)$ be the
probability at the critical temperature of such a crossing. One defines
similarly a vertical crossing and the associated probability $\pi_v(V,H)$.
It is a well-known fact from percolation theory that, on the triangular
lattice, a horizontal crossing on plus spins (or open sites)
exist if and only if there is no vertical crossing on minus spins
(closed sites). (One can convince oneself easily of this
simple fact using a drawing.) Since plus and minus spins are equiprobable,
this observation implies $\pi_h(V,H)+\pi_v(V,H)=1$. In this paper we will 
ultimately be interested in the limit $\pi_h(r)=\lim_{V,H\rightarrow \infty, r=
\sqrt3 V/2H}
\pi_h(V,H)$ and similarly for $\pi_v$. The infinite lattice limit of the previous
relation is known as the duality relation: $\pi_h(r)+\pi_v(r)=1$. If $\pi_h$
and $\pi_v$ are rotational invariants, the latter relation can be written as
$\pi_h(r)+\pi_h(1/r)=1$. This duality relation will hold for other
(regular) lattices if the functions $\pi_h$ and $\pi_v$ are universal but
its discrete equivalent ($\pi_h(V,H)+\pi_v(V,H)=1$) hold strictly for neither
the square nor the hexagonal lattices.

The present paper discusses both a prediction for $\pi_h$ extending Cardy's
approach and precise measurements of the function $\pi_h$ for $40$ values
of its parameter $r$. The agreement will be seen to be excellent, that is,
perfect within statistical errors. The first section
covers the theoretical prediction, the second the details of the simulation
and the comparison of the data with the prediction.

\section{A theoretical prediction based\\ on conformal field theory}

\subsection{Cardy's prediction for percolation}

Cardy's prediction for $\pi_h$ for two-dimensional percolation proceeds
in two steps. He first identifies the probability $\pi_h$ with the
difference of two partition functions with boundary conditions for
the $1$-state Potts model. He then uses the 
conformal field theory at $c=0$ to
obtain an analytic expression for this difference. Here is a (very
rapid) presentation of these two steps.

The partition function $Z(q)$ of the $q$-state Potts model on a finite
rectangular domain is the sum over all configurations $\sigma$
of $e^{-\beta H(\sigma)}$ where $H(\sigma)=J\sum_{\langle x,y\rangle}
(1-\delta_{\sigma(x), \sigma(y)})$. The sum in $H(\sigma)$ runs over
immediate neighbor pairs $\langle x,y\rangle$. This can be rewritten as 
$Z(q)=\sum_R p^{B(R)}(1-p)^{B-B(R)}q^{N_c(R)}$ where $p=1-e^{-\beta J}$.
The sum is over all subsets $R$ of the set of edges of the lattice in
the rectangular domain. The integer $B$ counts the edges in the lattice,
$B(R)$ those in the subset $R$ and $N_c(R)$ the clusters
in $R$. If $q=1$ (the value for percolation), this sum is 1 as desired.
Let $Z_{\alpha\beta}$ be the partition function of the Potts model for
configurations whose spins on the left side of the rectangle are in
the state $\alpha\in\{1,2,\dots,q\}$, those on the right in the
state $\beta$, and the others free. Cardy's first crucial observation
is that $\pi_h=\left.(Z_{\alpha\alpha}-Z_{\alpha\beta})\right|_{q=1}$
where $\alpha\neq\beta$. (The difference, done for a ``generic'' $q$,
contains precisely the configurations that have a cluster intersecting the
left and right sides.) The problem of calculating $\pi_h$ is therefore
transformed into that of calculating partition functions.

The possibility of calculating partition functions on finite domains
with given boundary conditions also originates from works by Cardy (see for
example \cite{C2}).
In the case of $Z_{\alpha\beta}$, for example, the four sides of the rectangle
are submitted respectively to the boundary conditions $\alpha$, free,
$\beta$ and free. Cardy argues that such a partition function is proportional
to the $4$-point correlation function $\langle \phi(z_1) \phi(z_2) \phi(z_3)
\phi(z_4) \rangle$ in the conformal field theory associated to percolation
whose central charge is $c=0$. The $z_i\in\mathbb C$ are the vertices of the
rectangle in the complex plane and $\phi$ is the field that changes the
boundary conditions in this theory. For percolation, this field is 
identified to $\phi_{1,2}$ and it has conformal weight $h=0$, a necessary
condition for $\langle \phi(z_1) \phi(z_2) \phi(z_3)
\phi(z_4) \rangle$ to be scale invariant. (The indices on $\phi_{1,2}$
refer to the labels of Kac table. See \cite{dFMS}.) The rules to find
correlation functions in a conformal field theory are well-known and
with this identification between partition functions with boundary conditions
and $4$-point correlation functions, the problem of calculating $\pi_h$ amounts
to solving an ordinary differential equation.

Two obstacles appear in applying these ideas to the Ising model. First the
Ising model is the $q=2$ Potts model and the difference $(Z_{\alpha\alpha}-
Z_{\alpha\beta})|_{q=2}$ cannot be interpreted as the crossing probability
since the factor $q^{N_c(R)}$ in the sum gives different weights to the various
configurations that have a crossing from left to right. Second although 
the operator
for the Ising model that changes the boundary state from free to a given state
$\alpha$ is still $\phi_{1,2}$, as in percolation, its conformal weight
$h_{1,2}$ is now $\frac1{16}$ and its $4$-point correlation is not anymore 
invariant under a conformal mapping $z\rightarrow w$ but picks up the usual
jacobian factors: $\langle \phi(w_1) \phi(w_2)\phi(w_3)\phi(w_4)\rangle=
(\prod_i |w'(z_i)|^{-h_{1,2}})\langle \phi(z_1) \phi(z_2)\phi(z_3)
\phi(z_4)\rangle$. These prefactors (and those additional coming from
the presence of vertices along the boundary) seem to contradict the (strict)
conformal invariance observed by simulation in \cite{LLS}.

In \cite{LPPS} and \cite{LPS} the probability $\pi_{hv}$ of having simultaneous
horizontal and vertical crossings was also obtained numerically. Watts \cite{W} was
able to extend Cardy's argument to obtain a prediction that fits extremely well
the data. His work is of particular interest to us as he is able to write
$\pi_{hv}$ again as a partition function but with more general boundary
conditions. Whether a conformal boundary operator accomplishes
the change between these more complex boundary conditions is not clear.
Nonetheless the expression of $\pi_{hv}$ as a partition function allows
us to expect, Watts argues, that this probability is given by some
$4$-point correlation function. He seeks it in the $h=0$ sector.

\subsection{The differential equation for $\pi_h$}

Cardy's argument for percolation cannot be extended to the Ising model.
One can still hope to relate crossing probabilities like $\pi_h$ to
$4$-point correlation functions as Watts did for $\pi_{hv}$ of percolation.
If such a relationship exists, the choice can be narrowed to $4$-point
functions of the identity family as these are the only ones in the
$c=\frac12$ conformal field theory that are invariant under conformal
map $z\rightarrow w$ like $\pi_h(r)$. An obvious objection will be
that the $4$-point function of the primary field in the identity family
is identically $1$ (or a constant). For the calculation at hand it might well be
that this primary field must be interpreted as one whose 
correlation functions satisfy only one of the differential equations
corresponding to the two leading singular vectors. This milder 
requirement does not force the function to be a constant as will be seen
immediately.

The Verma module $V_{(c=\frac12,h=0)}$ of the Virasoro algebra has a maximal
proper submodule $M$ generated by two singular vectors, one at level $1$,
the other at level $6$. The first of these is $L_{-1}|0\rangle$ and the
corresponding differential equation implies that the $4$-point function
is a constant. We shall drop this requirement. The other singular
vector is
\begin{eqnarray*}
  \left({L_{{-1}}}^{6}-10\,{L_{{-1}}}^{4}L_{{-2}}+{\frac {43}{3}}\,{L_{{-1}}}^
{2}{L_{{-2}}}^{2}-{\frac {100}{27}}\,{L_{{-2}}}^{3}+{\frac {97}{2}}\,{
L_{{-1}}}^{3}L_{{-3}}\right.\\
-{\frac {337}{6}}\,L_{{-1}}L_{{-2}}L_{{-3}}+{
\frac {3185}{144}}\,{L_{{-3}}}^{2}-{\frac {381}{2}}\,{L_{{-1}}}^{2}L_{{-4}}+{\frac
  {1265}{18}}\,L_{{-2}}L_{{-4}}\\
\left. +{\frac {19309}{36}}\,L_{{-1}}L_{{-5}}-{\frac
    {9005}{12}}\,L_{{-6}}\right) \left|c={\textstyle{\frac{1}{2}}},h=0\right\rangle . 
\end{eqnarray*}
If $f(z)$ is the $4$-point function with $z=(z_1-z_2)(z_3-z_4)/(z_1-z_3)
(z_2-z_4)$, the differential equation is
\begin{equation}
  \label{eqdiff}
  \begin{split}
    {\frac {1}{72}}\,\left (1-2\,z\right )\left (686\,{z}^{2}\left (1-z
      \right )^{2}+73\,z\left (1-z\right )+25\right ){\frac {d}{dz}}f(z)&\\
    +{\frac {1}{144}}\,z\left (1-z\right )\left (25141\,{z}^{2}\left (1-z
      \right )^{2}-2986\,z\left (1-z\right )-171\right ){\frac {d^{2}}{d{z}^
        {2}}}f(z)&\\
    +\frac{1}{
      27}\,{z}^{2}\left (1-z\right )^{2}\left (1-2\,z\right )
    \left (208-3595\,z\left (1-z\right )\right ){\frac {d^{3}}{d{z}^{3}}}f
    (z)&\\
    +\frac{1}{6}\,{z}^{3}\left (1-z\right )^{3}\left (137-737\,z\left (1-z
      \right )\right ){\frac {d^{4}}{d{z}^{4}}}f(z)&\\
    +10\,\left (1-2\,z\right ){z}^{4}\left (1-z\right )^{4}{\frac
      {d^{5}}{d{z}^{5}}}f(z)&\\
    +{z}^{5}\left (1-z\right )^{5}{\frac {d^{6}}{d{z}^{6}}}f(z)&=0.  
  \end{split}
\end{equation}
This differential equation has three (regular) singular points at $0,1$
and $\infty$. It is invariant under any permutation of these three points.
(Invariance under $z\rightarrow 1-z$ is clear: invariance under
$z\rightarrow 1/z$ requires some work.) The exponents
at any of these points are $0$, $\frac16$
twice degenerate, $\frac12$, $\frac53$ and $\frac52$. The monodromy
matrices around the three singular points are similar due to the symmetry
of the equation but they cannot be diagonalized simultaneously.

The cross-ratio $z=(z_1-z_2)(z_3-z_4)/(z_1-z_3)
(z_2-z_4)$ is related to the aspect ratio $r=\sqrt3 V/2H$ of the rectangle. If
the four points $z_1, z_2, z_3$ and $z_4$ are chosen along the real axis
at $-\frac1k, -1, 1, \frac1k$, then 
\begin{equation*}
k=\frac{1-\sqrt z}{1+\sqrt z}.
\end{equation*}
A Schwarz-Christoffel transformation can be used to map the upper plane
onto a rectangle with the images of the $z_i$ at the vertices. The aspect
ratio is then given as 
\begin{equation*}
r=\frac{K(1-k^2)}{2K(k^2)}
\end{equation*}
where $K$ is the complete elliptic integral of the first kind. The very
short but wide rectangles ($r=\sqrt3 V/2H\rightarrow 0^+$) corresponds to $z\rightarrow
0^+$, the tall and narrow ($r\rightarrow+\infty$) to $z\rightarrow 1^-$ and
the square to $r=1, z=\frac12$. The function $r(z)$ has the property
$r(z)=r(\frac1z)$ and the symmetry $z\rightarrow \frac1z$ of the differential
equation is thus welcome. If $\pi_h$ does not depend on the relative angle
between the rectangle and the lattice, that is if $\pi_h$ is a rotational
invariant, then the duality relation implies $\pi_h(1)=\frac12$ and it can
be put in the form $(\frac12-\pi_h(r))=-(\frac12-\pi_h(\frac1r))$. Fortunately
the function $r(z)$ is such that $r(z)=1/r(1-z)$ and the duality simply
states that the function $f(z)=\frac12-\pi_h(r(z))$ is odd with respect to
the axis $z=\frac12$. An odd subspace of the solution space of the differential
equation (odd with respect with $z=\frac12$) exists due to the symmetry
$z\rightarrow 1-z$ and it is of dimension $3$. 

We explored several
paths to cast the solutions of eq.\ (\ref{eqdiff}) into analytically tractable forms.
One of them was to write the lhs of eq.\ (\ref{eqdiff}) as 
$\prod_{1\le i\le6}(z(1-z))^{a_i}\frac{d}{dz}$ like Watts did.
But there is no real solutions for the $a_i$'s in the present case. The most
natural path however is the screening operator method (\cite{DF}, see also
\cite{dFMS}). The pertinent field is $\phi_{2,3}$ (with Kac's labels)
and the 4-point correlation calls for three contour integrals
(a charge $Q_+Q_-^2$ must be added at infinity to assure
neutrality). Integral representations of six linearly independent
solutions can be obtained in this straightforward (but probably tedious)
way. The main problem is therefore whether there is sufficient physical
information on $f$ to fix the linear combination. As argued above the odd parity
of $f$ reduces the space of solutions to a three-dimensional subspace.
The condition $f(0)=-\frac12$ (i.e.\ $\pi_h(0)=1-\pi_h(1)=0$) is one
further linear constraint. The function $f(z)$ is monotone increasing
but this condition will restrict $f$ to an open set of the $3$-d
subspace rather than decrease the dimension. The numerical data presented below indicate that 
$\pi_h(z)\rightarrow z^{\frac16}$ as $z\rightarrow 0^+$.
This is rather striking in view of the two-fold degeneracy of the
exponent $\frac16$. Two solutions associated to this exponent can
be chosen to behave as $z^{\frac16}$ and $z^{\frac16}\log z$ for small
$z$ and the latter, if present in $f$, should dominate the former
whenever $z$ is close to $0$. 
The fact that it is not seen in the simulation
could be interpreted physically as a manifestation of the power
law behavior of critical correlation functions at short distance. Imposing that
the behavior in $z^{\frac16}\log z$ be absent of $f$ would add one linear
constraint. With all these constraints we would still be left with
a one-dimensional subspace in the space of odd solutions. We have not
found any further constraints to fix completely the function $f$. This
is why we resorted to a numerical fit (see Paragraph \ref{numeric}).

As we were trying to solve analytically the differential equation, Marc-Andr\'e
Lewis suggested to us to look for a solution of the form 
$ez^d {}_2F_1(a,b,c;z)$, with $a,b,c,d$ and $e$ constants, that would be
odd around $z=\frac12$ and reproduce the asymptotic behavior of the data.
Cardy's prediction for percolation is of this form and the suggestion is natural
in this sense. Such a function exists but it does not satisfy the differential
equation. However it follows so closely the data of \cite{LLS} (the worst
gap is $1$\%) that we decided to improve these measurements to answer the
question: which of the hypergeometric function or of the solution of the
differential equation, if any, describes the data.

\section{Improved measurements of $\pi_h$}

\subsection{Finite size effects and power law behavior}

The probabilities $\pi_h,\pi_v$ and $\pi_{hv}$ at $81$ values
of $r$ ($\in[0.136,7.351]$)
were measured in \cite{LLS} for the three regular lattices on rectangles
containing around $40000$ sites. For these sizes, departure from the
duality relation is small but still noticeable for the square and the
hexagonal lattices. No attempt was made there to use various sizes in order to 
reduce finite-size effects. For the new runs to be presented here we
chose to concentrate on the triangular lattice and use various sizes
to approximate the function $\pi_h$ (and the other two, $\pi_v$ and $\pi_{hv}$) in the
limit when the mesh goes to zero.

A power law for the finite-size behavior of critical data is an accepted
hypothesis and our measurements rest upon it. It states that, for sufficiently
large size,
\begin{equation*}
|\pi_h(V,H)-\pi_h(r)| \approx a V^\beta
\end{equation*}
with $\beta$ a negative constant and $r=\sqrt3 V/2H$. We shall use several linear
sizes of the form $V=2^iV_0$ and $H=2^iH_0$. Writing $\pi_h(i)$ for
$\pi_h(2^iV_0,2^iH_0)$ and supposing that $\pi_h(i)$ is decreasing,
we can write $\pi_h(i)-\pi_h(i+1)\approx aV_0^\beta 2^{i\beta}(1-2^\beta)$ and
therefore express $\pi_h(r)$ as $\pi_h(i)-aV_0^\beta2^{i\beta}$. To determine the
constants $a$ and $\beta$ requires at least $3$ rectangle sizes as only
the differences $(\pi_h(i)-\pi_h(i+1))$ can be used.

\begin{table}[htbp]
  \begin{center}
    \leavevmode
    \begin{tabular}{|r|c|c||r|r|r|r|r|}\hline
      $r$ & $V_0$ & $H_0$ & $\hat\pi_h(0)$ & $\hat\pi_h(1)$ & $\hat\pi_h(2)$ &
      $\hat\pi_h(3)$ & $\hat\pi_h(4)$ \\
      \hline
      0.1443 & 4 & 24 & $0.029265|21$ & $0.025583|20$ & $0.023806|19$ & $0.022949|19$ & $0.022538|18$ \\ 
      \hline
      0.9897 & 16 & 14 & $0.502797|46$ & $0.499753|63$ & $0.498280|63$ & $0.497645|63$ & $0.497321|60$\\
      \hline
      6.928 & 32 & 4 & $0.978259|17$ & $0.977944|15$ & $0.977865|13$ & $0.977834|16$ & $0.977859|18$  \\ 
      \hline
    \end{tabular}
    \caption{$\hat \pi_h$ for three aspect ratios $r$ and five sizes. }
    \label{tab:pihinfini1}
  \end{center}
\end{table}

What are the right rectangle sizes and what is the required precision on
each $\hat\pi_h(i)$? For the two extreme rectangles that we are planning
to measure, $r_{\text{wide}}=\frac{\sqrt3}2\frac{V_0}{H_0}=
\frac{\sqrt3}2\frac4{24}\approx0.1443$ and
$r_{\text{tall}}=\frac{\sqrt3}2\frac{32}4\approx6.928$, and for a rectangle
close to a square $r_{\text{sq}}=\frac{\sqrt3}2\frac{16}{14}\approx 0.9897$,
we obtained $\pi_h(i)$, $i=0,1,2,3,4$. Each increment corresponds to an
increase by a factor of 2 of the linear size. For example $\pi_h(0)$ 
was measured on a rectangle of $4\times 24$ sites and $\pi_h(4)$ on
$64\times 384$ sites for the wide rectangle. The results appear in Table
\ref{tab:pihinfini1}. The samples were large, at least $250\times 10^6$
configurations. The digits
after the vertical bar gives the statistical error on the digits just
before; for example, 
the first element in the table ($0.029269|21$) means
that $\hat\pi_h(0)$ is $0.029269$ with the
95\%-confidence interval being $[0.029248, 0.029290]$. The differences between
$\hat\pi(3)$ and $\hat\pi(4)$ are however small. 
In fact the monotonicity of $\hat\pi_h(i)$ for $r_{\text{tall}}=
6.928$ is broken for $i=4$ even though the error bars do allow for
the power law to hold.
Larger samples would definitely be required
for the large lattices. Fortunately the precision on the measurements
for $i=0,1,2$ and the fact that the power law seems to hold for very small
sizes (4 sites in one direction!) allow for good estimates of $\pi_h(r)$ 
without these larger lattices. Table \ref{tab:pihinfini2} shows estimates
of $\pi_h$ for $r_{\text{wide}}$, $r_{\text{tall}}$ and $r_{\text{sq}}$
using the power law hypothesis and a subset of the measurements of
Table \ref{tab:pihinfini1}. The notation $i$--$j$ means that 
$\hat\pi_h(i),\hat\pi_h(i+1),\dots, \hat\pi_h(j)$ were used to obtain 
$\hat\pi_h(r)$. Using only the three smallest lattices, the three largest or the
five ones lead to estimates $\hat\pi_h(r)$ that differ by less than $4$
units on the fourth significant digits. We therefore decided to use only
three sizes for each $r$ considered and choose the pairs $(V_0,H_0)$
in such a way that $V_0, H_0\ge 4$ and that $V_0H_0\ge 96$. All samples
were larger or equal to $10^8$. Tables \ref{tab:resultatspih} 
gives the results for $\pi_h$, $\pi_v$ and $\pi_{hv}$ at
$40$ values of $r\in[0.1443, 6.928]$.

\begin{table}[htbp]
  \begin{center}
    \leavevmode
    \begin{tabular}{|r||r|r|r|r|r|r|}\hline
      $r$ & 0--2 & 0--3 & 0--4 & 1--3 & 1--4 & 2--4\\ \hline
      0.1443 & 0.02215 & 0.02215 & 0.02216 & 0.02215 & 0.02216 & 0.02216 \\ \hline
      0.9897 & 0.4969 & 0.4971 & 0.4970 & 0.4972 & 0.4970 & 0.4970 \\ \hline
      6.928 & 0.9778 & 0.9778 & 0.9778 & 0.9778 & 0.9778 & 0.9777 \\ \hline
    \end{tabular}
    \caption{Estimates $\hat\pi_h$ using subsets of available data.}
    \label{tab:pihinfini2}
  \end{center}
\end{table}

There is always, on a finite lattice, the problem of determining
$r$ from the numbers $V$ and $H$. The two simplest choices are the
aspect ratios of the smallest or the largest rectangles that include the
sites considered and only those. Even though it is not a natural choice,
the aspect ratio of the tallest and narrowest rectangle would be another
convention. The method we have used to determine $\pi_h(r)$ overcomes
this imprecision due to convention. For any convention the aspect
ratio for a rectangular subset of the triangular lattice will be
$\frac{\sqrt3}2(V+\Delta_V)/(H+\Delta_H)$ with $\Delta_V$ and $\Delta_H$
dependent on the convention but independent of $V$ and $H$. The ratio
$r$ at which $\pi_h$ is measured is therefore $\lim_{i\rightarrow
\infty} \frac{\sqrt3}2(2^i V_0+\Delta_V)/(2^iH_0+\Delta_H)=
\frac{\sqrt3}2 V_0/H_0$, independent of $\Delta_V$ and $\Delta_H$, 
that is independent of the convention. This is another advantage of using
several lattices for a given $r$.

Even though $\pi_h$ (and $\pi_v$) is invariant under rotation,
finite-size effects are not. The differences between 
$\hat\pi_h(0)=0.02927$,
$\hat\pi_h(1)=0.02558$ and
$\hat\pi_h(2)=0.02381$
for $r=0.1443$ are much bigger than those between 
$\hat\pi_v(0)=0.02174$,
$\hat\pi_v(1)=0.02205$ and
$\hat\pi_v(2)=0.02214$
for $1/r=1/6.928=0.1443$. The statistical errors on $\hat\pi_h(0.1443)$
and $\hat\pi_v(6.928)$ are therefore different even if both numbers
turn out to be very close ($0.02215$ and $0.02218$). In the worst
cases the $95$\%-confidence interval amounts to less than $2$ units
on the third significant digit (e.g.\ $\hat\pi_h(0.1443)=0.02215\pm
0.00015$). At the center of the range of $r$ the error on $\hat\pi_h(r)$
decreases to $4$ units on the fourth digit and it is even smaller for 
large $r$. Over the whole range it is smaller than $4\times 10^{-4}$
for both $\pi_h$ and $\pi_v$.

\begin{figure}
\begin{center}\leavevmode
\includegraphics[bb = 70 200 560 600,clip,width = 10cm]{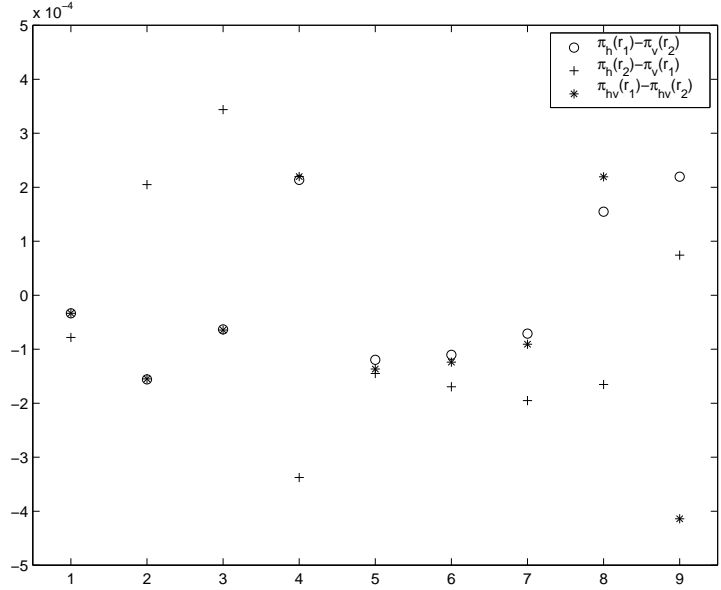}
\end{center}
\caption{Differences $\hat\pi_h(r_a)-\hat\pi_v(r_b)$ and
$\hat\pi_{hv}(r_a)-\hat\pi_{hv}(r_b)$ for several pairs
$(r_a, r_b=r_a^{-1})$.\label{fig:dua}}
\end{figure}

The improvement upon previous measurements found in \cite{LLS} can be
checked easily. Among the 40 values of $r$ used there are nine pairs
$((V_a,H_a),(V_b,H_b))$ such that 
\begin{equation*}
r_a=\frac{\sqrt3}2\frac{V_a}{H_a}=\left(\frac{\sqrt3}2\frac{V_b}{H_b}
\right)^{-1}=\frac1{r_b}.
\end{equation*}
The pairs $(r_a,1/r_a)$ are those corresponding to the following
lines of Table \ref{tab:resultatspih}: $(1, 40), \allowbreak
(5, 36),\allowbreak (7, 34), \allowbreak
(8, 33),\allowbreak (10, 31), \allowbreak
(11, 30), \allowbreak(12, 29), \allowbreak(18, 23), 
\allowbreak(20, 21)$.
The measurements for these pairs should satisfy $\hat\pi_h(r_a)=
\hat\pi_v(r_b)$, $\hat\pi_v(r_a)=\hat\pi_h(r_b)$ and $\hat\pi_{hv}(r_a)
=\hat\pi_{hv}(r_b)$ within statistical errors. This turns out to be
the case. (See Figure \ref{fig:dua}.)
There are in total 27 independent comparisons. Their relative
errors is always less than $2\times 10^{-3}$. The largest occur when the
quantities $\hat\pi$ being compared are themselves very small,
like $\hat\pi_h(\frac{\sqrt3}2\frac16)=0.02215$ and $\hat\pi_v(
\frac{\sqrt3}28)=0.02218$. In all cases the absolute value of
these differences are less than $5\times 10^{-4}$. Not only
are these variations small, they are of both signs. This fact is 
a further indication that the power law hypothesis provides a very good
approximation. Suppose indeed that, at the sizes used, correction
terms are required: $|\pi_h(i)-\pi_h(r)|\approx aV^\beta(1+\frac bV+\dots)$.
These new terms would lead to a systematic error that is not seen here.

\subsection{A prediction for $\pi_h$ for the critical Ising model}\label{numeric}

The first easy comparison between the theory developed in the first Section
and the new data lies in the asymptotic behavior of $\pi_h$ as $r$
approaches $0$ and $+\infty$. If $\pi_h$ is a solution of the differential
equation (\ref{eqdiff}), then $\log\pi_h(r)\rightarrow -\lambda\pi/r$
as $r\rightarrow 0$ and $\log(1-\pi_h(r))\rightarrow -\lambda\pi r$
as $r\rightarrow +\infty$ with $\lambda$ one of the exponents. Using the
ten extreme values of $r$ on each side of the measured interval,
we obtain for $\pi_h$
\begin{align*}
\log \hat\pi_h(r)&\underset{r\rightarrow 0}{\longrightarrow} -0.16648 \pi \frac1r\\
\log (1-\hat\pi_h(r))&\underset{r\rightarrow\infty}{\longrightarrow} -0.16657\pi r.\\
\end{align*}
The slopes obtained using the data for $\pi_v$ are $0.16647\pi$ and $0.16654\pi$.
These numbers are very close to the exponent $\frac16$, the smallest non-vanishing
exponent of the differential equation (\ref{eqdiff}). This is
remarkable! Clearly $\lambda$ is to be interpreted as a critical
exponent of the Ising model. But none of the usual exponents of the Ising model
contains the prime number $3$ in their denominators and scaling laws involve
only products and integral linear combinations of these exponents. If this new critical
exponent can be deduced from the usual ones, it will not be by traditional
scaling laws.

\begin{figure}
\begin{center}\leavevmode
\includegraphics[bb = 60 200 550 600,clip,width = 10cm]{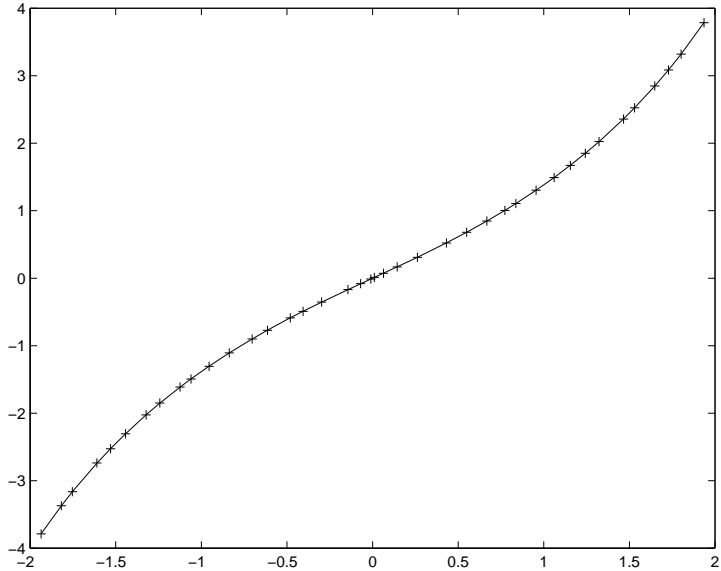}
\end{center}
\caption{The prediction $\log \pi_h/(1-\pi_h)$ as a function of $\log r$ together
with the 40 measurements $\hat\pi_h(r_i)$.\label{fig:pih}}
\end{figure}

The second test is to obtain a solution of the differential equation
(\ref{eqdiff}) that describes the data. Let $f_i,i=0,1,\dots,5$, be
a basis of solutions for
equation (\ref{eqdiff}) defined by their behavior at $z=\frac12$:
$f_i^{(j)}(z=\frac12)=\delta_{ij}$. Since $f(z)=\frac12-\pi_h(r(z))$
is odd with respect to $z=\frac12$, it lies in the subspace 
of functions of the form $af_1+bf_3+cf_5$. We determine the constants
$a, b$ and $c$ by requiring that $L=(\sum_i (\frac12-\hat\pi_h(r_i)
-f(z(r_i)))^2)^{\frac12}$
be minimum. The sum is over the 40 data. The three solutions $f_1,
f_3$ and $f_5$ were obtained numerically. 
Both {\scshape Matlab} 
and Mathematica 
give similar fits. These softwares have internal parameters controlling
the required accuracy of the integration. These parameters
can be pushed to a point where stronger
requirement does not lead to any significant improvement on the minimum of
$L$. Figure \ref{fig:pih} has been drawn using the values $a,b$ and $c$ 
obtained for control parameters beyond this point. The largest among the
differences $|\frac12-\hat\pi_h(r_i)-f(z(r_i))|, i=1,\dots, 40$, is
$3.6\times 10^{-4}$, smaller than the statistical error, and the
standard deviation is $1.5\times 10^{-4}$. Similar results are obtained for
$\pi_v$. The agreement is therefore excellent. As a comparison it is
instructive to redo Cardy's calculation testing his prediction for
percolation. Since the publication of \cite{Cardyperco}, better data
were obtained for percolation by sites on a square lattice
for $81$ rectangles with at least $10^6$ sites \cite{LPS}. The 
samples contained over $10^6$ configurations. For these, the statistical
errors are approximately $10^{-4}$ at the extremities of the
interval ($r\in[0.142, 7.067]$) and $10^{-3}$ in the middle. (At the
extremities these data are therefore more precise than the present ones for
the Ising model and at the center they are less precise.)
The largest departure
from duality is $4\times 10^{-4}$, almost exactly what is seen in
Figure \ref{fig:dua} for the present data. Comparing his prediction
$\pi_h^{\text{perco}}(r(z))=3\Gamma(\frac23)z^{\frac13}{}_2F_1(
\frac13,\frac23,\frac43,z)/\Gamma(\frac13)^2$ with the data
leads to $\max_{1\le i\le81}|\pi_h^{\text{perco}}(r(z_i))-\hat\pi_h(r_i)|= 
7.8\times 10^{-4}$ and to a standard deviation of $4.2\times 10^{-5}$.
These results are similar to those just reported.

We mentioned earlier the possibility of describing the data
with a hypergeometric function. The function 
\begin{equation*}
g(z)=\frac{6\Gamma(\frac13)}{\Gamma(\frac16)^2} z^{\frac16}{}_2F_1(
{\textstyle{\frac16}},{\textstyle{\frac56}},{\textstyle{\frac76}};z)
\end{equation*}
is odd around $z=\frac12$, when shifted by $\frac12$,
and behaves as $z^{\frac16}$ and $(1-z)^{\frac16}$
when $z\rightarrow 0^+$ and $z\rightarrow 1^-$. Several data are now
more than $2\times 10^{-3}$ apart from the corresponding values of $g$, 
a gap barely visible on a figure, but clearly out
of any reasonable confidence interval. We may accept the solution
of the differential equation as an analytic prediction for $\pi_h$ but
we must reject the function $g(z)$.

Measurements were also made of the probability $\pi_{hv}$. As its behavior
at $z=0$ and $z=1$ is also described by the exponent $\frac16$, one may
hope that the even subspace (around $z=\frac12$) of the differential
equation may contain a solution matching the data. This is not the case.
The best fit in this subspace lies up to $6\times 10^{-3}$ away from
the data, an unacceptable gap. (This disagreement
has an advantage. It shows that the success of the fit for $\pi_h$ is
not a consequence of the large freedom that a three-parameter
fit gives.) Watts used the third singular vector to describe $\pi_{hv}$
for percolation. In the Verma module $V_{c=0,h=0}$, this vector is at level
five, leading
to a differential equation of order $5$. However the third one in 
$V_{c=\frac12,h=0}$ is at level 11 and the associated operator
$P(z,\frac{d\ }{dz})$ is a polynomial
of order $11$ in $\frac{d\ }{dz}$. It can be cast as
\begin{align*}
P(z,\frac{d\ }{dz})=&\sum_{1\le i\le 11,\  i \text{\ odd}}((z(1-z))^{i-1}
p_i(z(1-z))\frac{d^i\ }{dz^i} \\
& +  \sum_{2\le i\le 10,\ i \text{\ even}}(1-2z)((z(1-z))^{i-1}
p_i(z(1-z))\frac{d^i\ }{dz^i}
\end{align*}
with
\begin{align*} 
p_1(u) & = -{\textstyle{\frac {1}{81}}}\,u\left (
-484-20465\,u+120702\,u^2+134456\,u^3+326536\,u^4\right )\\
p_2(u) & = -{\textstyle{\frac {1}{81}}}\left (-484-65267\,u
+555942\,{u}^{2}
+2442422\,u^3+9038782\,{u}^{4} \right )\\
p_3(u) & = {\textstyle{\frac {1} 
{162}}}\left (44318-1002950\,u-5132553\,u^2-12317152\,u^3+245463307\,u^4 
\right )\\ 
p_4(u) & = {\textstyle{\frac {1}{486}}}\left (-412839-8111249 
\,u-42237350\,u^2+ 
749236363\,u^3\right )\\ 
p_5(u) & = {\textstyle{-{\frac {1}{1944}}}}\left ( 
3492203+28198986\,u-1347384726\,{u}^{2}
+5106251212\,u^3\right )\\ 
p_6(u) & = {\textstyle{-{\frac {7}{1944}}}}\left (368339-27634444\,u+
153070553\,u^2\right )\\ 
p_7(u) & = {\textstyle{\frac {11}{1296}}}
\left (554551-9066926\,u+29200567\,{u}^{2}\right)\\ 
p_8(u) & = {\textstyle{\frac {11}{54}}}\left (- 
11842+75835\,u\right )\\ 
p_9(u) & = {\textstyle{-{\frac {11}{18}}}}\left (-757+3457\,u\right )\\ 
p_{10}(u) & = {\textstyle{-{\frac {110}{3 
}}}}\\ 
p_{11}(u) & = 1.
\end{align*} 
The differential equation $P(z,\frac{d\ }{dz})f(z)=0$
is again symmetric under any permutation of the
three regular singular points $0, 1$ and $\infty$.
The exponents (with their degeneracies) are $0 (2), \frac16 (2),
\frac12 (1), 1 (1), \frac53 (2), \frac52 (1), \frac{14}3 (1)$
and $6 (1)$.
The even subspace of the differential
equation is of dimension 6 and the fit 
will therefore contain 6 parameters.
Using the data of the last column of Table \ref{tab:resultatspih}
and numerical integration of the differential equation, we obtain
a best fit (in
the same sense as the one used for $\pi_h$) that has a largest
difference of $2.7\times 10^{-4}$ and a standard deviation of
$1.3\times 10^{-4}$. These are excellent results well within
the experimental windows. However we have tried to fit similarly
the function $h(z)= \kappa z^{\frac16}(1-z)^{\frac16}$ with the
constant $\kappa$ chosen such that $h(z=\frac12)=\hat\pi_{hv}(z=\frac12)$.
While this function {\em is not} a solution,
the 6-dimensional subspace of (numerical) even solutions contains
a solution $\tilde h$
that approaches it extremely well (namely
$\max_{z\in[0,1]}|\tilde h(z)-h(z)|\approx 10^{-6}$). Consequently
the large dimension of the even subspace allows for functions that
are not solution to be fitted within statistical errors and
the fit for $\pi_{hv}$
is much less convincing than the one
above for $\pi_h$ or than Watts' prediction.

\section*{Acknowledments} 
 
It is a pleasure to thank R.P.\ Langlands, M.-A.\ Lewis 
and G.\ Watts for helpful
discussions, A.\ Bourlioux for introducing us to the control of
errors in integrating differential equations and H.\ Pinson and
Ph.\ Zaugg for a careful reading of the manuscript.
 
E.\ L.\ gratefully acknowledges a fellowship from the NSERC Canada  
Scholarships Program and Y.\ S.-A.\ support 
from NSERC (Canada) and FCAR (Qu\'ebec).

\begin{table}[htbp]
\begin{center}
  \leavevmode
  \begin{tabular}{|c|c|c|r|r|r|}\hline
    $r$ & $V_0$ & $H_0$ & $\hat\pi_h$ & $\hat\pi_v$ & $\hat\pi_{hv}$ \\ \hline
      0.1443 & 4 & 24 & 0.02215 & 0.9779 & 0.02215  \\ \hline 
      0.1624 & 6 & 32 & 0.03321 & 0.9668 & 0.03321  \\ \hline 
      0.1732 & 6 & 30 & 0.04057 & 0.9594 & 0.04057  \\ \hline 
      0.1999 & 6 & 26 & 0.06082 & 0.9392 & 0.06082  \\ \hline 
      0.2165 & 6 & 24 & 0.07400 & 0.9257 & 0.07400  \\ \hline 
      0.2362 & 6 & 22 & 0.09062 & 0.9093 & 0.09061  \\ \hline 
      0.2665 & 8 & 26 & 0.1166 & 0.8830 & 0.1166  \\ \hline 
      0.2887 & 6 & 18 & 0.1359 & 0.8647 & 0.1359  \\ \hline 
      0.3248 & 6 & 16 & 0.1662 & 0.8341 & 0.1661  \\ \hline 
      0.3464 & 8 & 20 & 0.1834 & 0.8164 & 0.1831  \\ \hline 
      0.3849 & 8 & 18 & 0.2133 & 0.7865 & 0.2127  \\ \hline 
      0.4330 & 8 & 16 & 0.2483 & 0.7518 & 0.2468  \\ \hline 
      0.4949 & 8 & 14 & 0.2891 & 0.7111 & 0.2852  \\ \hline 
      0.5413 & 10 & 16 & 0.3163 & 0.6833 & 0.3096  \\ \hline 
      0.6186 & 10 & 14 & 0.3572 & 0.6429 & 0.3436  \\ \hline 
      0.6662 & 10 & 13 & 0.3797 & 0.6202 & 0.3600  \\ \hline 
      0.7423 & 12 & 14 & 0.4124 & 0.5878 & 0.3804  \\ \hline 
      0.8660 & 10 & 10 & 0.4580 & 0.5421 & 0.3997  \\ \hline 
      0.9326 & 14 & 13 & 0.4797 & 0.5203 & 0.4042  \\ \hline 
      0.9897 & 16 & 14 & 0.4969 & 0.5028 & 0.4059  \\ \hline 
      1.010  & 14 & 12 & 0.5029 & 0.4973 & 0.4058  \\ \hline 
      1.066  & 16 & 13 & 0.5183 & 0.4812 & 0.4043  \\ \hline 
      1.155  & 16 & 12 & 0.5419 & 0.4578 & 0.3996  \\ \hline 
      1.299  & 12 & 8 & 0.5767 & 0.4232 & 0.3860  \\ \hline 
      1.540  & 16 & 9 & 0.6277 & 0.3717 & 0.3544  \\ \hline 
      1.732  & 16 & 8 & 0.6639 & 0.3364 & 0.3269  \\ \hline 
      1.949  & 18 & 8 & 0.6999 & 0.3001 & 0.2952  \\ \hline 
      2.165  & 20 & 8 & 0.7318 & 0.2679 & 0.2655  \\ \hline 
      2.309  & 16 & 6 & 0.7516 & 0.2484 & 0.2469  \\ \hline 
      2.598  & 18 & 6 & 0.7864 & 0.2135 & 0.2128  \\ \hline 
      2.887  & 20 & 6 & 0.8163 & 0.1835 & 0.1833  \\ \hline 
      3.175  & 22 & 6 & 0.8418 & 0.1580 & 0.1579  \\ \hline 
      3.464  & 24 & 6 & 0.8643 & 0.1357 & 0.1357  \\ \hline 
      3.753  & 26 & 6 & 0.8834 & 0.1166 & 0.1166  \\ \hline 
      4.330  & 30 & 6 & 0.9137 & 0.08623 & 0.08623  \\ \hline 
      4.619  & 32 & 6 & 0.9259 & 0.07416 & 0.07416  \\ \hline 
      5.196  & 24 & 4 & 0.9452 & 0.05489 & 0.05489  \\ \hline 
      5.629  & 26 & 4 & 0.9563 & 0.04370 & 0.04370  \\ \hline 
      6.062  & 28 & 4 & 0.9651 & 0.03489 & 0.03489  \\ \hline 
      6.928  & 32 & 4 & 0.9778 & 0.02218 & 0.02218  \\ \hline
  \end{tabular}
  \caption{The measurements $\hat\pi_h$, $\hat\pi_v$ and $\hat\pi_{hv}$.}
  \label{tab:resultatspih}
\end{center}
\end{table}

\section*{List of captions}

\noindent Table \ref{tab:pihinfini1}: 
$\hat \pi_h$ for three aspect ratios $r$ and five sizes.

\noindent Table \ref{tab:pihinfini2}: 
Estimates $\hat\pi_h$ using subsets of available data.

\noindent Table \ref{tab:resultatspih}:
The measurements $\hat\pi_h$, $\hat\pi_v$ and $\hat\pi_{hv}$.

\noindent Figure \ref{fig:dua}: Differences $\hat\pi_h(r_a)-\hat\pi_v(r_b)$ and
$\hat\pi_{hv}(r_a)-\hat\pi_{hv}(r_b)$ for several pairs
$(r_a, r_b=r_a^{-1})$.

\noindent Figure \ref{fig:pih}: The prediction $\log \pi_h/(1-\pi_h)$ as a function of $\log r$ together
with the 40 measurements $\hat\pi_h(r_i)$.

\end{document}